\documentstyle[aps]{revtex}
\input{psfig.tex}
\begin{document}
\draft

\twocolumn[\hsize\textwidth\columnwidth\hsize\csname
@twocolumnfalse\endcsname

\title{
Two--Electron Atoms in Short Intense Laser Pulses
}
\author{Armin Scrinzi$^1$ and Bernard Piraux$^2$}
\address{$^1$ Institut f\"{u}r Theoretische Physik, Universit\"{a}t
Innsbruck,
Technikerstrasse 25,
A--6020 Innsbruck,  AUSTRIA}
\address{$^2$Institut de Physique,
Universit\'e Catholique de Louvain,
2, Chemin du Cyclotron,
B-1348 Louvain--la--Neuve, BELGIUM}

\maketitle

\begin{abstract}
We discuss a method of solving the
time dependent Schr\"odinger equation for atoms with
two active electrons in a strong laser field,
which we used in a
previous paper
{[}  A. Scrinzi and B. Piraux, Phys. Rev. A {\bf 56}, R13 (1997) {]}
to calculate ionization, double excitation and harmonic
generation in Helium
by short laser pulses.
The method employs complex scaling and an expansion in
an explicitly correlated basis.
Convergence of the calculations is documented and
error estimates are provided.
The results for Helium at peak intensities up to
 $10^{15} W/cm^2$ and wave length $248\,nm$ are accurate
to at least 10 \%.
Similarly accurate calculations are presented for
electron  detachment and  double excitation
of the negative hydrogen ion.
\end{abstract}

\pacs{PACS numbers: 32.80.Rm, 32.80.Fb, 31.20.Di}

\vskip1pc]
\narrowtext

\section{Introduction}
Several programs are being pursued that aim at the description of
three--dimensional two-- or multi-electron atoms in strong laser fields
\cite{burke90,burke97,parker96,zhang95}. The common motivation for these
efforts is to obtain quantitative results for excitation,
ionization, and generation of harmonics by laser pulses at intensities,
where more than one electron participates in the process.
The various approaches emphasize different aspects of the problem.

The first fully correlated three--dimensional calculations for
two--electron atoms in non--perturbative laser fields were made for
constant laser intensity by the R--matrix Floquet method \cite{burke90}.
Results have been published on $H^-$ and $H\!e$ \cite{purvis93}
and $Mg$ \cite{gebarowski97}.
The advantage of the method is that it
can be applied to multi--electron atoms, where
existing atomic structure programs can be used.
At large intensities, many angular momenta and Floquet blocks
are required and very large systems of equations have to be solved.
Recently an adaptation of the R--matrix method to solve the
time--dependent Schr\"odinger equation has been proposed \cite{burke97},
which maintains the applicability to general atoms, but may be less 
plagued
by expansion size problems.

The approach of Ref.~\cite{parker96} puts a strong emphasis on
two--electron correlation in $He$--like atoms at the expense of
abandoning the realistic description of atomic structure.
The method solves the time--dependent Schr\"odinger equation
on a grid for the radial electron coordinates and with
an expansion in single particle spherical harmonics for the
angular degrees of freedom. By visualization of the wave function,
in particular the process of direct double ionization could be
studied. The implementation is adjusted to a massively parallel 
computer,
but still the grid size and the length of the
multipole expansion of the inter--electron
potential is limited by computer resources.

The method of Ref.~\cite{zhang95} describes a two--electron
wave function by an expansion in numerical single--electron wave 
functions
that are calculated in a finite box. This provides a realistic
representation of atomic structure and  allows to adjust the wave 
function
to the parameter range to be investigated. For example, when 
photoelectron
spectra are to be extracted, continuous wave functions can be densely
placed in range of electron energies of interest.
Results have been published on $Mg$ \cite{zhang96}.

Here we present in detail the method employed for the calculation
of excitation, ionization, and harmonic generation in $H\!e$
published in a previous paper \cite{scrinzi97}.
Our purpose is to provide converged {\it ab initio} calculations
for realistic laser parameters with special emphasis on electron
correlation. We use an expansion in explicitly correlated two--electron
basis functions and complex scaling \cite{reinhardt}. The range of
application is similar to that of Ref.~\cite{zhang95}. The most 
important
difference is our use of an explicitly correlated basis, which gives
a very accurate description of atomic structure including
doubly excited states with a relatively short expansion.
The second crucial ingredient of our method is complex scaling
which, as we will show,  gives a simple
implementation of strictly outgoing wave boundary conditions
by an $L^2$ method. The penalty of the method is the loss
of a direct physical
interpretation of the continuous spectrum of the complex
scaled operator.
While this may not be a fundamental limitation of the complex scaling
method,
it does at present limit our results to total ionization,
double excitation, and harmonic generation.

Compared to Ref.~\cite{scrinzi97} we extend the calculations
for Helium to higher laser intensities up to $10^{15}W/cm^2$ at
the laser wave length of $248\,nm$.
The foundation of the error estimates given in Ref.~\cite{scrinzi97}
is presented and discussed in detail.
By improvements of the basis the accuracy of the harmonic spectra
calculation could be enhanced to about 10\%.
We supplement
the results by laser detachment and double excitation of $H^-$.

\section{Computational method}

The Schr\"odinger equation of a two--electron atom exposed to a laser 
field
described in velocity gauge with the dipole approximation  is
\begin{equation} 
i\frac{d}{dt}\Psi( \vec{r}_1, \vec{r}_2;t)=
\left[H_0
\label{e:schroedinger}
+\frac{i}{c}\vec{A}(t)\cdot( \vec{\nabla}_1+ \vec{\nabla}_2)
\right]
\Psi( \vec{r}_1, \vec{r}_2;t)
\end{equation} 
with the atomic Hamiltonian
\begin{equation} 
H_0=
-\frac{1}{2}(\Delta_1+\Delta_2)
-\frac{Z}{r_1}-\frac{Z}{r_2}+\frac{1}{| \vec{r}_1- \vec{r}_2|},
\label{e:hamatom}
\end{equation} 
where $ \vec{r}_1$ and $ \vec{r}_2$ denote the electron coordinates 
measured from
the nucleus and $\Delta_i$ and $ \vec{\nabla}_i$ are the corresponding
 Laplace  and gradient operators. The nuclear charge is $Z=2$ for Helium
and $Z=1$ for the negative hydrogen ion. Atomic units are used unless
stated otherwise.

The vector potential of a  linearly polarized laser pulse is given by
\begin{equation} 
\vec{A}(t)=
h(t) \sin( \omega t)
\left(0,0,A_0\right),
\end{equation} 
where we employed $\cos^2$ and Gaussian shaped envelopes
\begin{eqnarray*} 
h_{\cos^2}(t)&=&[\cos(\pi t/T)]^2\\
h_{\rm Gauss}(t)&=&\exp[-(2t/T)^2]
\end{eqnarray*} 
with the pulse duration $T$.

The calculations were made in velocity gauge, since we found much
better convergence than in length gauge, which is in
agreement with previous experience and with theoretical arguments
\cite{cormier96}. A calculation at intensity $10^{14}W/cm^2$ and
frequencies of $ \omega=0.4$ and 0.6 was repeated in
length gauge and gave the same results for ionization and single
excitation. Results for double excitation could not be converged
in length gauge.

Eq.~(\ref{e:schroedinger}) is  a $6+1$--dimensional equation, which
can be reduced to $5+1$ dimensions because of cylindrical symmetry, when
the laser is linearly polarized. Due to the high dimensionality
only a very limited range of the phase space can be numerically
represented and one needs to carefully control the restrictions
imposed on this space.
The restrictions consist of basis set truncation and the
boundary conditions at large distance.
We first discuss the boundary conditions.

\subsection{Absorption of outgoing flux}

Ionization means that a finite portion of the wave function
moves away to arbitrarily large distances without further
contributing to the dynamics of the system.
In a finite space one needs to absorb this outgoing flux
at the boundary of the space to avoid unphysical reflections.
Common procedures are the use of a complex potential
at large distances \cite{kulander87,burke97} or some form
of mask function \cite{parker96}. A more systematic control
of the asymptotic boundary conditions has been proposed
in Ref.~\cite{zhang96a}.
Ideally, one admits only outgoing waves at large distances.
However, outgoing wave boundary conditions are difficult to define
in the presence of a dipole field, which ranges to arbitrarily
large distances. In any case,
correctly imposed outgoing wave boundary conditions are
energy dependent, which is, in general, quite difficult
to implement computationally. An additional complication is
that the resulting Hamiltonian
is non--selfadjoint (the norm of the wave function
on the finite space is not conserved)
and it has non--orthogonal eigenfunctions.
This may cause problems for computational implementations that
rely on the orthogonality of the eigenfunctions of the Hamiltonian.

For calculations with only one active electron, which are effectively
$2+1$--dimensional, one can usually make the space large enough
such that the boundary conditions are of secondary importance.
In our case a more stringent method of absorbing outgoing
flux is required.
Such a method is complex scaling \cite{reinhardt,simon82}.
It consists in analytically continuing the
Hamiltonian by multiplying the real coordinates by
a complex number
\begin{equation}\label{e:complex0} 
H( \vec{r}_1, \vec{r}_2;t)\to H_{\theta}=H(e^{i\theta} 
\vec{r}_1,e^{i\theta} \vec{r}_2;t),
\end{equation} 
where the scaling angle $\theta$ is real and positive.
For the time--independent Schr\"odinger equation, the mathematical
theory of complex scaling is well established.
The new Hamiltonian $H_{\theta}$
has the same bound state spectrum as $H$,
 while the continuous spectrum is rotated by the
angle $-2\theta$ around
the ionization thresholds into the lower half plane of complex energies.
This separates the continua starting from different ionization 
thresholds.
In the wedge--shaped area between the real axis and the
rotated continua doubly excited states appear as square
integrable eigenfunctions
with complex eigenvalues, whose imaginary parts give 1/2 of the
autoionization widths.
In an exact calculation, the values of bound state and
resonance energies do not depend on the scaling angle $ \theta$.
The method is being widely applied. For multi-photon
physics it is used to calculate ionization
rates and AC--Stark shifts of hydrogen--like systems  by the
Floquet method and in time--dependent calculations
for hydrogen--like systems \cite{pont91,huens93}.

There is no complete mathematical theory for the
application of complex scaling
to time dependent problems, only partial results for the time
evolution of bound and resonance states were found \cite{buchleitner95}.
In the appendix we argue that the restriction of the complex
scaled Schr\"odinger equation
\begin{equation}\label{e:schrcomp} 
i\frac{d}{dt}\Psi_{ \theta}( \vec{r}_1, \vec{r}_2;t)=H_{ \theta}\Psi_{ 
\theta}( \vec{r}_1, \vec{r}_2;t)
\end{equation} 
to the space of square integrable functions is equivalent
the unscaled equation with the constraint of strictly outgoing
wave boundary conditions.
The outgoing wave solution at the coordinates $( \vec{r}_1, \vec{r}_2)$
is obtained by evaluating $\Psi_{ \theta}$
at the back--scaled arguments $(e^{-i \theta} \vec{r}_1,e^{-i \theta} 
\vec{r}_2)$.
To establish this equivalence we need
to assume far reaching analyticity properties of the solution
$\Psi_{ \theta}( \vec{r}_1, \vec{r}_2;t)$,
which are difficult to prove in practice.

Regardless of this mathematical problem,
the method has been successfully employed in
time--dependent calculations \cite{huens93} and
its validity could be verified numerically \cite{mccurdy89}.
For hydrogen one can approach the limit $ \theta\to0$,
i.e. directly compare with the usual Schr\"odinger equation.
It was found that the  projections on bound states
and the expectation value of the dipole
\begin{equation}\label{e:dipole} 
\vec{d}(t)=
\langle \Psi_{ \theta}(e^{-i \theta} \vec{r},t) 
|  \vec{r} | \Psi_{ \theta}(e^{-i \theta} \vec{r},t)  \rangle
\end{equation} 
do not depend on the scaling angle $ \theta$. The advantage of the
complex scaled solution is that due to the absence of
reflections a much
shorter expansion of the wave function can be used when $ \theta\not=0$.
For the two--electron system basis size requirements exclude
very small scaling angles, but we found stable results for
the excited state populations and for
$\vec{d}(t)$ in
the range of $0.12 \alt \theta \alt0.28$ (see below).

The harmonic spectrum is obtained by Fourier
transforming the acceleration of the dipole $\ddot{\vec{d}}$.
The total ionization yield is defined as
\begin{equation}\label{e:pionize0} 
Y_{\rm ion}=1-\sum_{i}
| \langle \Phi_{i}( \vec{r}_1, \vec{r}_2)
|  \Psi_{ \theta}(e^{-i \theta} \vec{r}_1, \vec{r}_2;t=\infty)  
\rangle|^2,
\end{equation} 
where $\Phi_{i}$ is the $i$--th
bound state function calculated with the real Hamiltonian $H_{ 
\theta=0}$.
We use the computationally more convenient formula
\begin{equation}\label{e:pionize1} 
Y_{\rm ion}=1-\sum_{i}
| \langle \Phi^*_{i, \theta}( \vec{r}_1, \vec{r}_2)
|  \Psi_{ \theta}( \vec{r}_1, \vec{r}_2;t=\infty)  \rangle|^2.
\end{equation} 
where $\Phi_{i, \theta}$ is the bound state eigenfunction of
the complex scaled atomic Hamiltonian
\begin{equation} 
H_{0, \theta}\Phi_{i, \theta}=E_i\Phi_{i, \theta}.
\end{equation} 
Note the extra complex conjugation on the
left hand function, i.e. in the integral the {\em unconjugated}
function is used.
Eqs.~(\ref{e:pionize0}) and (\ref{e:pionize1}) are equivalent because of
the analyticity of both, $\Phi_{i, \theta}$ and $\Psi_{ \theta}$, and 
since
for $ \theta=0$ $\Phi_{i, \theta=0}$ is real up to an overall phase 
(see appendix).

The population of a doubly excited state $ \alpha$ is determined as
\begin{equation} 
P_{ \alpha}=| \langle \Phi^*_{ \alpha, \theta} | \Psi_{ \theta}  
\rangle|^2.
\end{equation} 
This equation does not have an unscaled analogue, since
the resonance wave function
$\Phi_{ \alpha, \theta}$ seizes to be square integrable, when
$ \theta$ approaches 0.

\subsection{Basis set expansion}
\label{s:basis}

We approximate the solution of the complex scaled
Schr\"odinger equation by expanding
$\Psi_{ \theta}$ in a Hylleraas-like explicitly correlated basis
\begin{eqnarray} 
\lefteqn{
\Psi_{ \theta}( \vec{r}_1, \vec{r}_2;t)=P_1
\sum_{L=0}^{L_{\rm max}}
\sum_{ l=0}^{L}
G_{L l}( \vec{r}_1, \vec{r}_2)}&&
\nonumber\\
\quad&&\times \sum_s \sum_{k=0}^{k_s} \sum_{m=0}^{m_s} \sum_{n=0}^{n_s}
c^{L l}_{kmn;s}(t) r_1^{k} r_2^{m} r_{12}^n
e^{- \alpha_s r_1 - \beta_s r_2}
\label{e:expansion}
\end{eqnarray} 
The operator $P_1$ projects on the singlet
states and $r_{12}:=| \vec{r}_1- \vec{r}_2|$.
The  two--electron angular factors $G_{L l}$ for
total angular momentum $L$ and $z$--component $L_z=0$ are
\begin{equation} 
G_{L l}=r_1^{ l}r_2^{L- l}
\sum_m C^{L,0}_{ l,m;L- l,-m}
Y^{ l}_m( \hat{r}_1)Y^{L- l}_{-m}( \hat{r}_2).
\end{equation} 
The $C^{L,0}_{ l,m;L- l,-m}$ are Clebsch--Gordan coefficients and
$Y^{l}_m$ are spherical harmonics.
Note that for each $L$ there are only $L+1$
angular functions
$G_{L, l}$. The major part of angular correlation,
which in the usual atomic
physics basis requires a large number of combinations
of single--electron angular momenta $ l$ and $L- l$, is here contained
in the inter--electron coordinate $r_{12}$.

The expansion (\ref{e:expansion})
is known to be formally complete \cite{king67}
and it converges rapidly for bound states
of the three--body Coulomb system. In Ref.~\cite{kono84}
a further significant improvement of the basis was achieved by
selecting the combination of powers by the rule
\begin{equation}\label{e:constraint} 
k+m+n+|k-m|(1-\delta_{0n}) \leq p_{s}.
\end{equation} 
This constraint can be understood as follows:
The range of space covered by a basis function in the direction
of $r_1$ and $r_2$ is roughly $k/ \alpha_s$ and
$m/ \beta_s$, respectively.
When $ \alpha_s\sim \beta_s$ and  $|k-m|$ becomes large,
the electrons remain far from each other and correlation,
which is mostly contained in the coordinate $r_{12}$,
becomes small. One therefore needs fewer functions with
$r_{12}$--dependence, when  $|k-m|$ is large.
The constraint leads to an important reduction in the expansion
size without deteriorating the accuracy of bound and doubly
excited state energies.

In Refs.~\cite{kono84,sims88} for each state of Helium
two sets of exponents were used, one
describing the known asymptotic behavior of the bound state wave 
function
by selecting $ \alpha_1=Z$ and $ \beta_1=\sqrt{-Z^2-2E}$ and a second 
one
describing correlation by exponents $ \alpha_2= \beta_2$, which
were optimized to obtain the best bound state energy $E$.
In our case we needed to describe many states, including
strongly correlated doubly excited states, within the same basis set.
Therefore we used several different sets of exponents for each
$L$ and $ l$. As an example, the exponents and the powers used for
$L=2$ in the major part of the calculations
are given in Table~\ref{t:basisl2}.
The first group of exponents is adjusted to describe the
singly excited states and single electron continuum
of the configuration type
$(1s,n'd)$, the middle group is for symmetrically doubly excited
states and higher continua
of the form $(np,n'p)$ and the last group is for
states $(nd,n's)$ with single electron quantum numbers
$n,n'=1,2,3$. The particle--exchanged configurations
are automatically included by the exchange symmetrization of
the basis functions.

It is important to observe that in
velocity gauge double excitation must be
included into the basis for a correct representation of
the wave function, even when
no real double excitation occurs. The reason is that the gauge
transformation
\begin{equation} 
\Psi( \vec{r}_1, \vec{r}_2;t)\to
e^{i\vec{A}(t)\cdot( \vec{r}_1+ \vec{r}_2)}\Psi( \vec{r}_1, \vec{r}_2;t)
\end{equation} 
equally affects both coordinates and thus introduces virtual double
excitation.

Our basis functions
$G_{L l}( \vec{r}_1, \vec{r}_2) r_1^{k} r_2^{m} r_{12}^ne^{- \alpha_s 
r_1 - \beta_s r_2}$
are strictly real, such that the phase as well as the $ 
\theta$--dependence
of $\Psi_{ \theta}( \vec{r}_1, \vec{r}_2;t)$ is entirely contained in 
the
expansion coefficients.
This means that
the expansion coefficients of our complex scaled initial state
$\Phi_{ \theta,1S}( \vec{r}_1, \vec{r}_2)$
are dependent on $ \theta$ and for each $ \theta$ and
a different system of equations with different initial condition
has to be solved.

We can interpret the same fact in terms of the
back--scaled solution $\Psi_{ \theta}(e^{-i \theta} \vec{r}_1,e^{-i 
\theta} \vec{r}_2;t)$,
which approximates the solution of the normal Schr\"odinger equation 
with
outgoing--wave boundary conditions.
The expansion functions for the outgoing--wave solution
are then
\[
e^{-i(L+k+m+n) \theta}
G_{L l}( \vec{r}_1, \vec{r}_2) r_1^{k} r_2^{m} r_{12}^n
e^{-e^{-i \theta}( \alpha_s  r_1 + \beta_s r_2)},
\]
i.e. they strongly depend on $ \theta$.
By varying $ \theta$ we therefore vary the expansion functions
for the physical solution and in this way we obtain an estimate of the
basis set truncation error with respect to the radial
coordinates $r_1,\,r_2$ and $r_{12}$.

\subsection{Alternative basis sets}

The good performance of expansion (\ref{e:expansion}) for
$He$ is due to the fact that on the one hand it is very similar to the
usual atomic physics expansion in terms of products of
single--electron orbitals, which converges well for states,
where the two electrons remain spatially separated.
On the other hand
the explicit dependence on $r_{12}$ allows a good description
of the wave function at small inter--electronic distances,
which is particularly important when both electrons are in the
same shell, as in the ground state or in symmetrically excited
states.

We investigated several other expansions, which
seem offer technical advantages or which are particularly
suitable for specific states of $He$.

An implementation of arbitrary angular momentum for few--body
systems is given by Wigner's $D$--functions\cite{wigner,varshalovich89}.
At high
angular momenta that expansion allows a strong reduction
of the number of non--zero matrix elements in the calculation
\cite{scrinzi96}. The $D$--functions separate
the overall rotation of the system from internal degrees of
freedom. The overall rotation is defined as the rotation
between a body--fixed
coordinate system, determined by two vectors $\vec{a}$ and $\vec{b}$,
and the laboratory coordinates.
The $D$--functions carry indices $D^L_{mn}$, which designate the
total angular momentum $L$, the quantum number $m$ of rotation around
the lab--fixed $z$--axis, and the quantum number $n$ of the rotation
around $\vec{a}$.
Although there is a certain freedom
of choice for $\vec{a}$ and $\vec{b}$, they cannot be identified
with the electron coordinates $ \hat{r}_1$ and $ \hat{r}_2$, since
the definition of the $D$--functions is not symmetric
under exchange of $\vec{a}$ and $\vec{b}$.
In order to implement electron exchange symmetry, one can for example
use the Jacobi coordinates
\begin{eqnarray*} 
\vec{a}&=& \vec{r}_1+ \vec{r}_2\\
\vec{b}&=& \vec{r}_1- \vec{r}_2
\end{eqnarray*} 
as the body--fixed vectors.
With this choice,
 the subscripts of $D^L_{mn}$ refer to collective rotations
of the electrons. This is desirable for some highly correlated
states, like the Wannier states \cite{rost}. In the
unsymmetrically excited states that dominate the
wave function of an atom excited by a laser pulse,
one electron carries the major part of angular momentum (except for
symmetrization), which leads to poor convergence of the $D$--function
expansion.

Similar problems arise for an expansion with respect to
the perimetric coordinates
\begin{eqnarray*} 
u&=&-r_1+r_2+r_{12}\\
v&=&\quad r_1-r_2+r_{12}\\
w&=&\quad r_1+r_2-r_{12}.
\end{eqnarray*} 
While the interparticle coordinates $r_1,r_2,r_{12}$ are subject
to the triangular inequality $|r_1-r_2|\leq r_{12} \leq r_1+r_2$,
the perimetric coordinates
$u,\,v$, and $w$ each vary independently in $[0,\infty)$.
This simplifies the calculation of integrals and allows to
find expansion functions, where the operator matrices become sparse.
Like with  the Jacobi coordinates,
the unsymmetrically excited states, where the two
electrons move largely independently, are not efficiently described
by such an expansion, since $u,\,v$, and $w$ each contain both
coordinates $r_1$ and $r_2$. It also appears difficult to find a
constraint like (\ref{e:constraint}) to cut down on the basis size.

Finally, we explored an expansion with respect to the
coordinates $r_1,\,r_2,$ and $\cos \theta_{12}:= \hat{r}_1\cdot 
\hat{r}_2$.
This is very similar to
an expansion in single electron orbitals. The main advantage is that
the calculation of matrix elements becomes simple. However, the
expansion length is generally larger than with the
explicitly correlated basis and bound state accuracies
beyond $10^{-4}\,a.u.$ become extremely hard to achieve 
\cite{scrinzi96}.

\subsection{Numerics and computation}
\label{s:numerics}

The expansion (\ref{e:expansion}) is notoriously numerically
difficult. The main reason is that the metrical matrix
\begin{equation} 
S_{ij}:= \langle i| j  \rangle
\end{equation} 
for the basis functions
\begin{equation} 
|i \rangle=G_{L l}( \vec{r}_1, \vec{r}_2)r_1^{k_i}r_2^{m_i}r_{12}^{n_i}
e^{- \alpha_s r_1- \beta_s r_2}
\end{equation} 
rapidly becomes ill--conditioned with increasing powers 
$k_i,\,m_i,\,n_i$.
The situation is further aggravated by our use of several different sets
of exponents $ \alpha_s,\, \beta_s$ for the same $L$, which makes the 
basis
formally overcomplete.
We were able to control these problems by performing accurate 
integrations,
by appropriately normalizing the basis,
and by removing near singular values from
the metrical matrix.
Still, at high angular momenta we needed to resort to
Fortran REAL*16 ($\approx32$ decimal digits)
accuracy in the calculation of the matrix elements.

We first express
the angular factors in the form
\begin{equation} 
G( \vec{r}_1, \vec{r}_2)=\sum_j g_j r_1^{k_j}r_2^{m_j}r_{12}^{n_j}
(\cos \theta_1)^{ \lambda_j}(\cos \theta_2)^{\mu_j},
\end{equation} 
where $\cos \theta_i$ is the cosine of $ \vec{r}_i$ with the $z$--axis.
The determination of the expansion coefficients $g_j$ is straightforward
but a little cumbersome. Except for an insignificant overall
factor, the $g_j$ are rational numbers with not too large denominators,
which allowed us to compute them numerically and ensure that they were
accurate to all digits of arithmetic precision.

For the angular integration we change variables to
$\cos \theta_1,\,\varphi_1,\,\cos \theta_{12}$ and $\varphi_{12}$, where
$\varphi_1$ is the azimuthal angle of $ \vec{r}_1$,
$\cos \theta_{12}= \hat{r}_1\cdot \hat{r}_2$, and $\varphi_{12}$ is the 
azimuthal
angle of $ \vec{r}_2$ with respect to the axis $ \vec{r}_1$.
Integrations can then be performed over all angular variables except
for $\cos \theta_{12}$, which is expressed as
$\cos \theta_{12}=(r_1^2+r_2^2-r_{12}^2)/(2r_1r_2)$.
The remaining three--dimensional integrals have the general form
\begin{equation}\label{e:integrals} 
\int dr_1dr_2dr_{12} r_1^k r_2^m r_{12}^n e^{- \alpha r_1- \beta r_2}.
\end{equation} 
The integrals are non--trivial, because $r_1,\,r_2$ and $r_{12}$
are connected by the triangular inequality. We compute them by
a strictly positive and therefore numerically
stable recurrence formula \cite{scrinzi92}.

Loss of accuracy is due to the expansions of $G_{L l}$
and $\cos \theta_{12}$, which at high $L$ become very lengthy
with up to several hundred terms. Beyond $L\approx5$ we needed
to use Fortran REAL*16 arithmetic, which provided stable
results up to $L\approx12$.

Even when no accuracy is lost in the matrix elements the metrical
matrix $S$ has in general a very poor condition number which quickly
exceeds regular machine precision ($\approx14$ decimal digits) or
even REAL*16 precision ($\approx32$ decimal digits).
This may lead to uncontrolled errors. We remove the
problem by first rescaling the basis such that the diagonal
elements of the metrical matrix are $S_{ii}\equiv1$, after which
the maximum size of the eigenvalues of $S$ was limited to
$ \alt 100$.
We diagonalize $S$ and remove eigenvectors with eigenvalues smaller
than a threshold $ \epsilon$. We found $ \epsilon=10^{-11}$ to be 
suitable for
Fortran REAL*8  arithmetic.
In REAL*16 we could use $ \epsilon=10^{-30}$, but results were
found to depend very weakly on the threshold.
The remaining eigenvectors $|\xi_i \rangle$ are normalized
with respect to $S$
\begin{equation} 
\langle \xi_i | S | \xi_j  \rangle =  \delta_{ij}. 
\end{equation} 
After transforming all matrices to the orthonormal
basis $\{|\xi_i \rangle\}$  precision can be lowered to REAL*8 to save 
on
storage and computation time.

\subsection{Time propagation}
\begin{figure}
\centerline{\psfig{figure=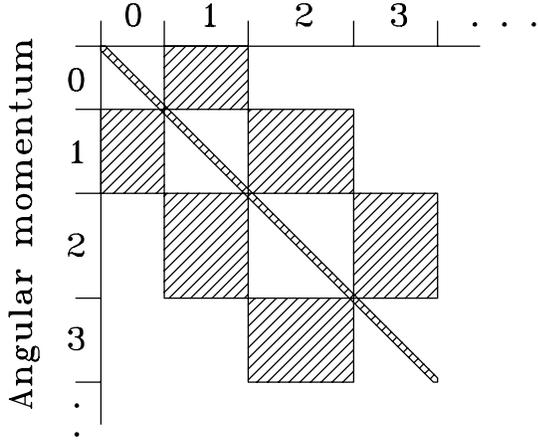,width=7cm,angle=90}}
\caption{\label{f:hamiltonian}
The Hamiltonian matrix in the atomic basis}
\end{figure}

For the time propagation we make one more transformation
to the ``atomic basis'' $\{|\eta_i \rangle\}$ that diagonalizes the 
complex
scaled atomic Hamiltonian
\begin{equation} 
\langle \eta_i | H_{0, \theta} |\eta_j \rangle=E_i \delta_{ij}, 
\end{equation} 
where
\begin{equation} 
H_{0, \theta}=
-\frac{e^{-2i \theta}}{2}(\Delta_1+\Delta_2)
-\frac{Ze^{-i \theta}}{r_1}-\frac{Ze^{-i \theta}}{r_2}
+\frac{e^{-i \theta}}{r_{12}}.
\end{equation} 
In dipole approximation the laser field
only connects angular momenta differing by $|L-L'|=1$.
The resulting structure of the overall Hamiltonian is depicted in
Fig.~\ref{f:hamiltonian}.

At the laser intensities and frequencies in our calculations the 
diagonal terms
still make the dominant contribution to the time evolution. Therefore
we time integrate in the ``interaction picture''
\begin{eqnarray} 
c(t)&=&a(t)+b(t)\\
i\frac{d}{dt}a(t)&=&\hat{H}_{0, \theta}a(t)\label{e:atom}\\
i\frac{d}{dt}b(t)&=&\left[\hat{H}_{ \theta}(t)-\hat{H}_{0, 
\theta}\right]a(t)
+\hat{H}_{ \theta}(t)b(t).
\label{e:interaction}
\end{eqnarray} 
Here $c(t)$ denotes the coefficient vector  and $\hat{H}_{0, \theta}$
and $\hat{H}_{ \theta}$ are the operator matrices with respect to the
$\{|\eta_i \rangle\}$ basis. For the decomposition of $c(t)$ into 
$a(t)+b(t)$
we set at the beginning of each time step
\[
a(t_0)=c(t_0),\qquad b(t_0)=0.
\]
Since $\hat{H}_{0, \theta}$ is diagonal in our basis, the
solution of Eq.~(\ref{e:atom})
is trivial. Eq.~(\ref{e:interaction}) is solved by
a seven stage 6th order explicit Runge--Kutta method
(Butcher's method, given in Ref.~\cite{book73}),
which we found to be more CPU time
efficient than lower order methods. There were no problems with 
numerical
stability, as we verified by comparing with lower order methods
 at a few parameter points.
The time step was automatically adapted by comparing every two 
integration
steps with a single double step size integration.
The typical number of time steps was about 400 per optical cycle, which
for our 7 stage method means about 3000 matrix--vector multiplies per 
cycle.
Computation times for the shorter pulses were about 1 hour on a
500 MHz DEC/Alpha work station.

\section{Results}

\subsection{Bound and doubly excited states}
Before we discuss the results of the time propagation, we want
to list the energies and widths of the most important bound
and doubly excited states of $He$ and $H^-$
as obtained with the above basis.
Table~\ref{t:energies1} compares the values of the first
few bound state energies of $He$ that we obtained with the basis
used in the time propagation with  reference values from literature.
Several of our values are lower than the variational upper bounds
from literature, but this does not indicate greater accuracy, since
due to complex scaling our values are not upper bounds.
Our accuracy is $\sim 10^{-8}a.u.$ for most energies
given.
The basis sizes are about 300 for each angular momentum.
The functions are counted
after removal of near--singular vectors from the basis
(cf. section~\ref{s:numerics}), which is the number relevant for the
time propagation. As an example, we listed in Table~\ref{t:basisl2}
above the basis set for $L=2$.
It has been shown that literature values can be exactly reproduced
with bases of size $\sim 150$ that are optimized for each state
\cite{kono84,sims88}.

Table~\ref{t:energies2} gives lowest few
doubly excited states of $He$ from the time--propagation.
The states are labelled by the
radial quantum numbers of the dominant single electron contributions
$n_1$ and $n_2$.
Accuracies
are of the order $10^{-4}\,a.u.$ for the energies and widths.
At the very small widths the relative accuracies can become poor.

Finally Table~\ref{t:energies3} summarizes the bound and doubly
excited state energies of $H^-$.
We present the energies as they appear in the time propagation
as well as values  obtained in larger
bases with state specifically adapted exponents $ \alpha_s$ and $ 
\beta_s$.
Judging from the convergence behavior,
we believe that our state--specific values are accurate to all
except possibly the last digit quoted.
One sees that some of them are more accurate than the literature
values quoted.
A special case is the
 shape resonance in $L=1$ just above the
$H(n=2)$ threshold. It requires a minimum scaling angle of
$\theta\approx0.25$
to become manifest as an isolated eigenvalue of the scaled Hamiltonian.
With the smaller angle used in the time propagation the resonance
state cannot be distinguished from the approximate continuous states
that surround it.

\subsection{Excitation and ionization of $He$}

\begin{figure}
\centerline{\psfig{figure=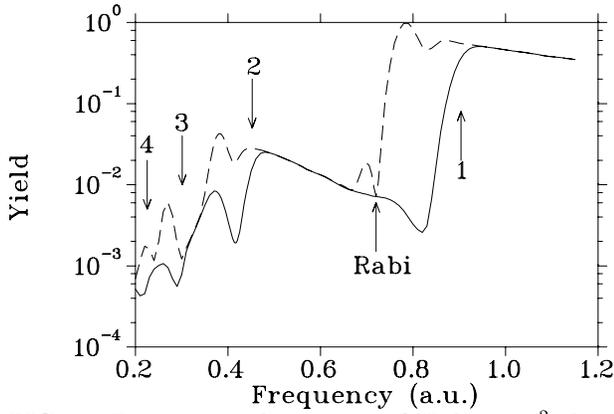,width=8cm,angle=90}}
\caption{\label{f:ionization}
Excitation and ionization of $He$ by a $\cos^2$--shaped pulse of
duration 3.8 fs and peak intensity $2.97\times10^{14}W/cm^2$ as
a function of frequency.
Solid line: ionization, dashed line: ionization plus bound state 
excitation.
The arrows labelled by $n=1,2,3,$ and 4 indicate
$n$--photon ionization thresholds. The dip at frequency 0.72 is due to
a Rabi oscillation.
}
\end{figure}

\begin{figure}
\centerline{\psfig{figure=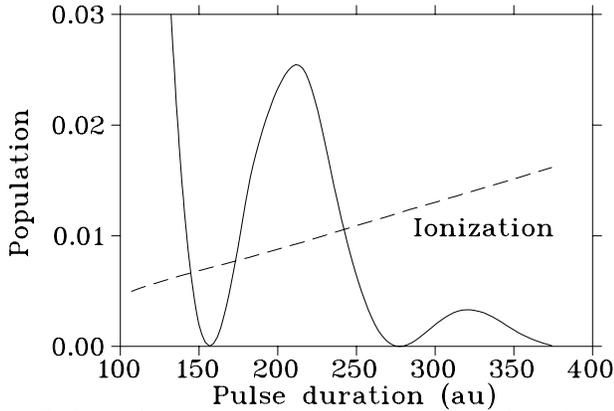,width=8cm,angle=90}}
\caption{\label{f:rabi}
Bound state excitation as a function of pulse duration at
frequency $ \omega=0.72$. Solid line: total population
of excited bound states. Dashed line: ionization.
The distance between the minima is roughly the Rabi period.
}
\end{figure}

\begin{figure}
\centerline{\psfig{figure=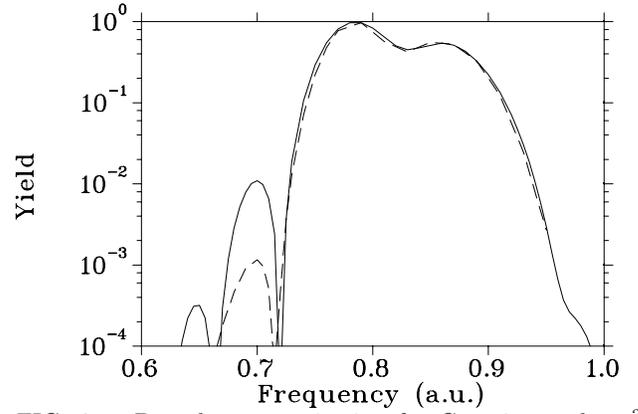,width=8cm,angle=90}}
\caption{\label{f:gauss}
Bound state excitation for Gaussian and $\cos^2$ pulses as
a function of frequency. The pulse energy is $0.5a.u.$ for
both shapes, the pulse widths are $T=157$ for $\cos^2$ pulses
(solid line) and
$T=92$ for Gaussian pulses (dashed line).
}
\end{figure}

Fig.~\ref{f:ionization} shows the probability of excitation and
ionization of $He$ by $\cos^2$--shaped pulses of duration
$T=157\,a.u.$ and peak intensity 
$I=0.00423\,a.u.=2.97\times10^{14}W/cm^2$.
Note that in Ref.~\cite{scrinzi97} the conversion to SI units was too 
small
by a factor of 2. The frequencies cover the range from just above the
five photon ionization threshold to well above
the single ionization threshold. Below each of the thresholds one
clearly distinguishes the enhancement of bound state excitation due
to resonances. The peaks below the one photon and two photon thresholds
are due to resonances with the lowest $P$--state (energy = $-2.12384$)
and the lowest excited $S$--state (energy $= -2.14597$), respectively.
The resonances below the lower thresholds are not well separated due
to the spectral width of the pulse.
The pronounced dip in the bound state excitation at $ \omega=0.72$
is due to a Rabi--like oscillation. When one increases the
pulse duration, the minimum disappears completely and reappears
at a pulse duration of $\sim 270$. A third minimum appears
at $T=380$ (Fig.~\ref{f:rabi}). The oscillation period roughly 
corresponds
to the Rabi period at the given parameters.
The effect strongly depends on the pulse shape.
Fig.~\ref{f:gauss} compares  excitation by $\cos^2$--pulses with
excitation by Gaussian pulses. The pulse energy is $0.5\,a.u.$ for
both pulse shapes and the widths of the vector potential
envelopes was $T=157$ for the $\cos^2$--pulse and $T=92$ for
the Gaussian pulse. With this choice the envelopes have the
same width at $1/e$ of the maximum for both pulse shapes.
One can see that the Rabi--like oscillations occur for both
shapes, but they are more pronounced with the $\cos^2$--pulses.

One can also distinguish resonant enhancement of ionization,
manifested by the coincidence of bound--state excitation peaks
with ionization peaks. Most pronounced are the peaks at
the three--photon resonance  with the lowest $P$--state at
frequency $ \omega=0.26$ and the two-photon resonance with the lowest
excited $S$--state at $ \omega=0.38$; another resonance with the lowest
$D$--state is hidden in the slope to the two--photon ionization
threshold. In spite of the massive bound state excitation
at the single--photon resonance at $ \omega=0.78$, weak coupling
of the $P$--state to the continuum leads only to a slight
bump in the ionization rate at that frequency.

The lowest frequency of 0.2 in the figures is somewhat above the
the popular frequency of 0.1837 of a  $248 nm$ wavelength laser.
In Table~\ref{t:248}
we give the ionization yield at wave length $248\,nm$ by a $\cos^2$ 
pulse of
duration $T=40\,{\rm cycles}=32.8\,fs$, which has a half width
of 20 optical cycles. The pulse parameters coincide with the
ones used in Ref.~\cite{charalambidis97} for the comparison
with experiment. Our results are systematically lower by about
50\% compared to Ref.~\cite{charalambidis97}.
We estimate that our results are converged to 10\% accuracy or
better (see below). Considering that we cover intensities
all the way into the onset of saturation, the agreement is
nevertheless satisfactory.

Recently another calculation for the same pulse has been published
\cite{mercouris97}. Those values sizably oscillate around ours
and also around the results
of Ref.~\cite{charalambidis97}. The disagreement
is particularly evident at the lower intensities, where convergence 
problems
should be smaller. It appears therefore that calculations of
Ref.~\cite{mercouris97}
remain relatively far from convergence.
A possible reason for the lack of convergence is the absence of
doubly excited states from the basis of Ref.~\cite{mercouris97},
although the velocity gauge used in that calculation
always introduces at least virtual
doubly excited states (cf. section \ref{s:basis}).

The experimental number given in Ref.~\cite{charalambidis97}
is at least one order of magnitude smaller than all theoretical results
and none of the calculations falls within the quite large
upper error margin of the experiment.

\begin{figure}
\centerline{\psfig{figure=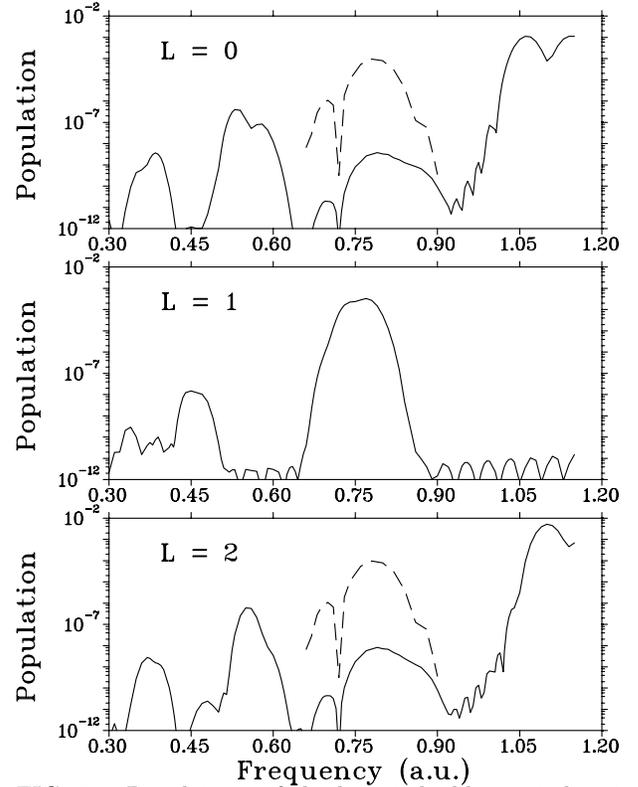,width=8cm,angle=0}}
\caption{\label{f:dxall}
Populations of the lowest doubly excited states with $L=0, 1,$
and 2 after the passage of a pulse with $T=3.8\,fs$ with
peak intensity $2.97\times10^{14}W/cm^2$. Dashed lines: population
of the lowest bound state with symmetry $^1P^o$ $\times 10^{-4}$.
}
\end{figure}

Ionization is predominantly a single electron effect and all
observations made above are qualitatively the same in a single
electron atom. Double excitation in turn is a genuine
two electron effect. Fig.~\ref{f:dxall} shows the population
of the lowest doubly excited states with angular momentum
$L=0,\,1,\,2,$ and 3 after passage of a $\cos^2$--shaped
pulse of duration $T=157$. The narrower peaks are due to multiphoton
resonances between the ground and the respective doubly excited states
with 2 to 6 photons. The population of the states with even $L$ are
enhanced in the range $ \omega=0.6\sim0.9$. This enhancement is almost
exactly proportional to the $^1P^o$ bound state population and
the ratio between the populations was found to be proportional to
the laser intensity. We therefore interpret it as a far off--resonant
single photon transition from the $P$ to the $S$ and $D$ states
brought about by the large band width of the pulse. As with
the Rabi oscillation of single excitation,  the pulse
shape is quite important, since with Gaussian pulses the
phenomenon nearly disappears. Similarly, longer pulse duration
suppresses the effect.

\subsection{Excitation and electron detachment from $H^-$}
The binding energy of the only bound state of $H^-$ is $0.02775\,a.u.$.
Therefore much lower laser intensities and frequencies lead to total
electron detachment from $H^-$.

\begin{figure}
\centerline{\psfig{figure=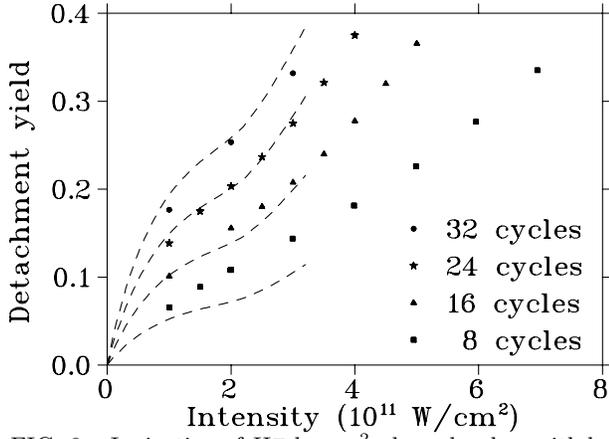,width=8cm,angle=90}}
\caption{\label{f:hm-om3}
Ionization of $H^-$ by $\cos^2$--shaped pulses with laser
frequency $ \omega=0.03$ as a function
of intensity for pulse durations $T$ = 8, 16, 24, and 32 optical cycles.
The  bend in the ionization yield
is due to the closure of single--photon ionization channel by the
AC--Stark shift and the transition to
two--photon ionization.
Dashed lines: time--integrated Floquet rates from Ref.~[5]
}
\end{figure}

Fig.~\ref{f:hm-om3} shows the photodetachment at a laser frequency of
$ \omega=0.03$ for intensities between $1\times10^{11}W/cm^2$ and
$8\times10^{11}W/cm^2$. At about $2\times10^{11}W/cm^2$ the single 
photon
ionization threshold rises above $ \omega=0.03$ due to the AC Stark
shift and roughly in the same intensity region two photon ionization
becomes dominant \cite{purvis93}.
In our time--dependent calculations we can distinguish a
bend in the detachment yield, which moves towards expected intensity
of  $2\times10^{11}W/cm^2$ with increasing pulse duration.
The dashed lines are obtained by integrating the
intensity dependent
detachment rates $ \Gamma(I)$ from Ref.~\cite{purvis93} with the pulse
shapes used in our calculations. Saturation effects
in the final detachment yield $Y(t=\infty)$
are included according to the equation
\begin{equation} 
Y(t)=\int_{-\infty}^t dt'\,  \Gamma[I(t')]\, [1-Y(t')].
\end{equation} 
As the instantaneous intensity we defined
\begin{equation} 
I(t):=[E_0 h(t)]^2/2,\qquad E_0= \omega A_0.
\end{equation} 
This definition neglects terms with the time--derivative
of the envelope, whose contributions do not visibly change the 
comparison.
The agreement for the two longer pulses is quite good, except that
the bend in the yield is somewhat more pronounced for the
integrated rates.
With the shorter pulses  the
integrated rate underestimates the true detachment.
Assuming the applicability of the rate concept also at the
short pulses, the spectral width of the pulse
could be sufficient to explain the higher yield:
with a broad pulse the closure of the
single photon ionization channel is moved to higher intensities,
thus effectively enhancing the yield. For a quantification of this
hypothesis rates $ \Gamma(I)$ for frequencies other than $ \omega=0.03$ 
would be
required.

\begin{figure}
\centerline{\psfig{figure=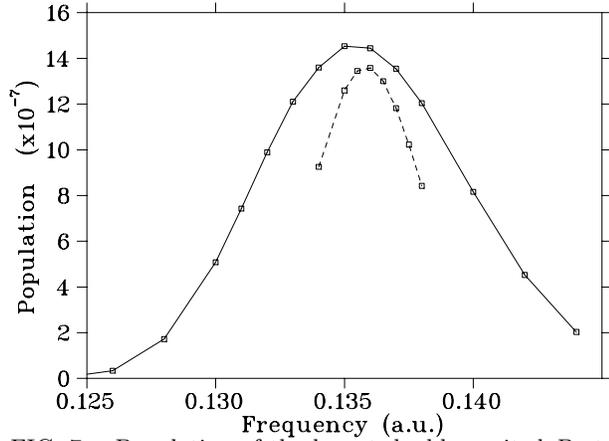,width=8cm,angle=90}}
\caption{\label{f:hm-dxp}
Population of the lowest doubly excited $P$--state
of $H^-$ by $\cos^2$--shaped pulses as a function of frequency.
Peak intensity is
$10^{13}W/cm^2$ and pulse duration duration $T=10$ optical cycles
(solid line) and $T=20$ (dashed line).
}
\end{figure}

For the double excitation of $H^-$ one needs frequencies in the
visible to low UV. Fig.~\ref{f:hm-dxp} shows the excitation of
the relatively long--lived lowest autoionizing $P$ state
around the  resonant frequency of $ \omega=0.134$ and intensity 
$10^{13}W/cm^2$.
The resonance is  shifted
from its field--free position by about $0.001$ for the
10 cycle pulses. Longer pulses of 20 cycles have a larger shift
of $\sim 0.002$. In the limit of a constant laser field
an AC--Stark shift of the ground state by $\sim 0.005$ is expected
by perturbatively extrapolating the data from Ref.~\cite{purvis93}

\subsection{Harmonic generation}
By Eq.~(\ref{e:dipole}) we calculate the expectation
value of the dipole $\vec{d}(t)$ as a function of time, from
which we obtain the acceleration of the dipole by numerical 
differentiation.
It would be more desirable to directly calculate the expectation
values of the acceleration of the dipole $\ddot{\vec{d}}(t)$,
but unfortunately in  the matrix elements of that operator
integrals of the form
(\ref{e:integrals}) with negative powers of $r_1$ or $r_2$ arise,
which can only be calculated with considerable extra
numerical effort (cf. \cite{scrinzi92}).

\begin{figure}
\centerline{\psfig{figure=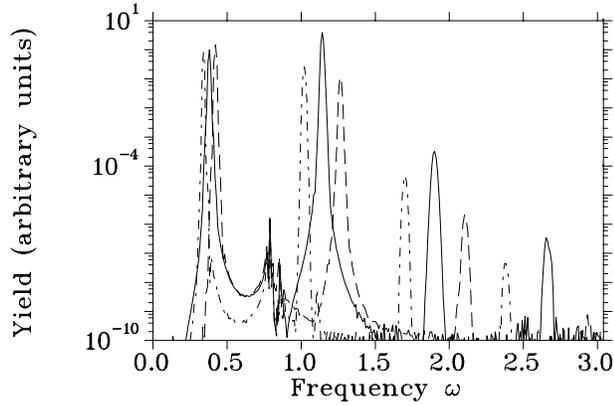,width=8cm,angle=90}}
\caption{\label{f:harm}
Harmonic generation on Helium with a $\cos^2$--pulse
of duration T = 40 optical cycles and peak intensity 
$2\times10^{14}W/cm^2$.
The fundamental frequencies are 0.34 (dot--dashed line), 0.38 (solid 
line)
and 0.42 a.u. (dashed line), respectively.
The frequency 0.38 is two--photon resonant with the lowest
excited $S$--state. The structure around
$ \omega=0.8$ originates from bound state excitations.
}
\end{figure}

The expectation values $\vec{d}(t)$ are sensitive to
the wave function at larger distances.
To obtain accurate results we therefore
added extra basis functions with smaller exponents to cover
a longer range in $r_1$ and $r_2$. By this enlargement of the basis we
were able to obtain satisfactory accuracies of $ \alt 10\%$ up to the
5th harmonic for frequencies in the range 0.34 to $0.42\,a.u.$ and
intensity $I=2\times10^{14}W/cm^2$.
The frequencies include a strong resonance with the lowest
excited $S$--state that greatly enhances harmonic generation and
leads to the dominance of the 3rd harmonic over the 1st.
Figure~\ref{f:harm} summarizes the results obtained
with a pulse duration of 40 optical cycles.
Harmonics up to order 7 can be distinguished and the resonant
enhancement at $ \omega=0.38$ is manifest.
For reference we include Table~\ref{t:harm} with the peak heights.

\subsection{Accuracy of the results}

The limited expansion length for the wave function introduces
the dominant error into our calculations. Now we study the
effect of the truncations
with respect to total angular momentum and radial basis
functions.

\begin{figure}
\centerline{\psfig{figure=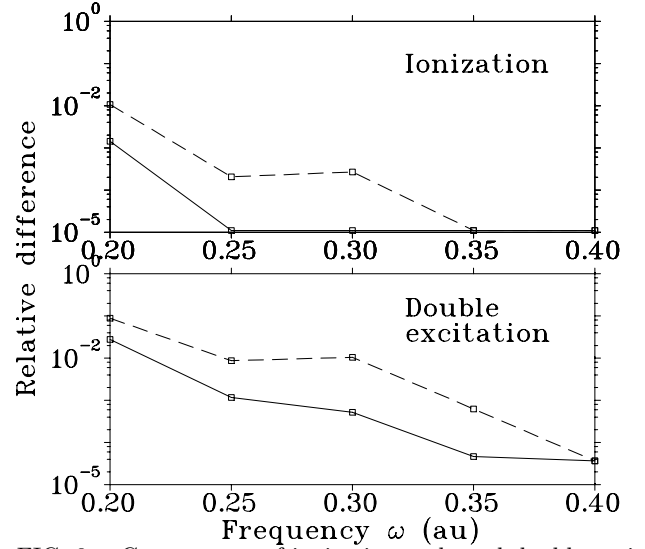,width=8cm,angle=90}}
\caption{\label{f:l-dep}
Convergence of ionization and total double excitation with the
maximum angular momentum $L_{\rm max}$. Dashed line: relative
difference between calculations with $L_{\rm max}=5$ and 7; solid
line: relative difference between  $L_{\rm max}=6$ and 7. Intensity
$=2.97\times 10^{14}W/cm^2$, pulse duration $T=157$.}
\end{figure}

Fig.~\ref{f:l-dep} compares the results of calculations with
$L_{\rm max}=5$, 6, and 7 at peak intensity
$2.97\times10^{14}W/cm^2$ and frequencies between
$0.2$ and $0.4$. Only at the lowest frequencies there
is a distinguishable effect on ionization with a relative
difference between the calculations of $ \alt1\%$.
Double excitation is more sensitive to angular
momentum truncation, since it is a much higher order process, but
still the relative error rises to only about $10\%$.
Assuming exponential convergence,
we can conclude that with $L_{\rm max}$ also for double excitation
the error due to angular momentum truncation
is $ \alt 1\%$ at the given parameters.

\begin{figure}
\centerline{\psfig{figure=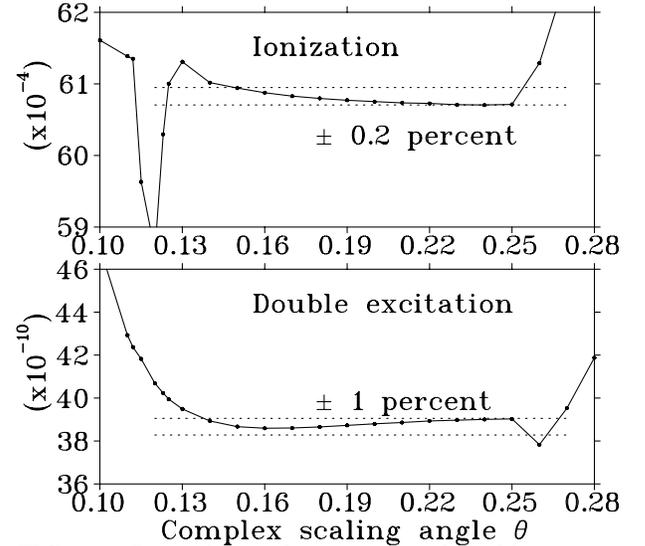,width=8cm,angle=90}}
\caption{\label{f:th-dep}
Dependence of ionization and double excitation on the complex scaling
angle $\theta$. The dotted lines indicate estimated errors.
Intensity
$=2.97\times 10^{14}W/cm^2$, frequency=0.4 and pulse duration $T=157$.
}
\end{figure}

Convergence of the expansion in the internal coordinates
$r_1,\,r_2,\,r_{12}$ is more difficult to investigate, since the
basis rapidly grows when one increases
the admissible powers of $r_1,\,r_2,\,r_{12}$.
Complex scaling provides an indirect estimate of the accuracy of the
internal expansion. For an infinite basis, the results are
independent of $ \theta$. Any dependence on $ \theta$ must therefore be
ascribed to the basis truncation. In practice, there is only
a limited range of $ \theta$ where the results vary little with
$ \theta$. When $ \theta$ is too small, the outgoing waves
are only weakly damped and one needs to describe a long
oscillatory tail in the wave function, for which an $L^2$--expansion
converges slowly.
When on the other hand $ \theta$ is too large
the complex scaled bound state wave functions have
increasingly oscillatory
character, which again is not well reproduced by the finite basis
(cf. appendix).
Fig.~\ref{f:th-dep} shows ionization and double
excitation obtained with different scaling angles at
laser frequency $0.39\,a.u.$.
One distinguishes
a range of $ \theta$ where the results are quite stable. The variation
inside this range is of the same size as the variation
when we increase the  number of basis functions by a factor $\sim 2$, 
which
supports our use of the variation with $ \theta$ for an accuracy 
estimate.
For the given parameters ionization varies less than 0.2\%.
The accuracies for double excitation are only slightly lower.

At frequency $ \omega=0.3$ the variation of ionization still remains
within the $1\%$ range, but double excitation
varies by about 10\%,
which indicates
that we approach the limits of numerically reliable results
for multiphoton double excitation.

\section{Summary and conclusions}
The method introduced in this paper allows the numerical integration
of the complete Schr\"odinger equation of a two--electron atom in a 
strong
laser pulse with realistic frequency, intensity, and duration.
This has been demonstrated on the systems of Helium and $H^-$ with
frequencies ranging from infrared to ultraviolet, intensities
up to $10^{15}W/cm^2$ and pulses as long as $160\,fs$.

The first important constituent of our method is an explicitly
correlated basis set expansion, which allows to reproduce the
atomic structure to essentially any desired accuracy at moderate 
expansion
length. This includes positions and widths of doubly excited states, 
where
we obtain accuracies that rival and in some cases exceed literature 
values.

The second ingredient of our method, complex scaling, was formally 
introduced
as a method of imposing strictly outgoing boundary conditions.
We cannot at present give a mathematically rigorous theory for
the use of complex scaling in a time--dependent calculation, but
we provide heuristic arguments and numerical evidence that it indeed
is equivalent to the regular Schr\"odinger equation with strictly
outgoing boundary conditions.
The technical advantage of complex scaling is that the expansion
length remains short.

Convergence was investigated for the whole range of parameters and
indicates accuracies between fractions of a percent at higher laser
frequencies and at least 10\% for the majority of the data. Only at the
seventh harmonic peak in our example and for double excitation at
lowest laser frequencies accuracies remain unsatisfactory.

The numerically most challenging combination of parameters was used
for Helium exposed to pulses of duration $\sim 32\,fs$ at the wave 
length
of $248\,nm$ and peak intensities up to $10^{15}W/cm^2$, where
we reach an accuracy of the ionization yield of about $10\%$.
A previous calculation for the same pulses \cite{charalambidis97}
qualitatively agrees with ours but exceeds our result by about
50\% at the highest intensity. A more recent calculation
\cite{mercouris97} deviates more strongly also at the lower intensities.

Another comparison with existing theoretical work could be
performed with a Floquet calculation for the electron detachment
from $H^-$. Quite satisfactory agreement was found for pulses of at 
least
16 optical cycles. Shorter pulses may not be expected to compare
well with a Floquet calculation for constant intensity.

We believe that our results should serve as a benchmark for future
calculations of two--electron systems. This refers to both, complete
two--electron calculations as well as model calculations. It may
be expected that for harmonic generation and ionization a 
single--electron
description will be found to be satisfactory in a range of parameters.

If necessary, our method allows extensions in several directions.
To extend  the range of accessible parameters is predominantly a
question of more computer power, although also technical modifications
in the calculation of the matrix elements are required to  control the
loss of accuracy.
A second obvious extension is the introduction of an electronic core
to model effective two--electron atoms like $Mg$.
Finally, the interpretation of the complex scaled wave function
adopted in this paper, which takes the back--scaled wave function
as an approximation to the regular solution with outgoing wave boundary
conditions, suggests that electron spectra can be determined.
Whether this is numerically feasible and practical remains to be
investigated.

\section*{Acknowledgement}
We wish to thank Robin Shakeshaft and Marcel Pont for numerous
fruitful discussions.
A.~S.\ acknowledges support by the APART program
of the Austrian Academy of Sciences and thanks for
the hospitality enjoyed during
several stays at the Universit\'e Catholique de Louvain.
B.~P.\ is chercheur qualifi\'e au
Fonds National de la Recherche Scientifique
of Belgium.

\appendix
\section{}
Here we deduce the relation between outgoing wave boundary
conditions and complex scaling.
We first assume that one can write the radial wave function in the form
\begin{equation} 
\Psi(r;t)=\int_{-\infty}^{\infty} dk\,c(k,t)\Phi_k(r),
\end{equation} 
where
the $\Phi_k$ have the asymptotic behavior $\sim e^{ikr}$,
and $\Psi$ solves the Schr\"odinger equation
\begin{equation}\label{e:schroe1} 
i\frac{d}{dt}\Psi(r;t)=H(r;t)\Psi(r;t).
\end{equation} 
For the sake of brevity we have omitted the part of
the expansion with square--integrable functions.
(For the Coulomb potential the asymptotic behavior is more
precisely $\sim \exp(ikr-i\ln 2|k|r/k)$.)
In terms of the expansion coefficients $c(k;t)$ Eq.~(\ref{e:schroe1})
can be written as
\begin{equation}\label{e:schroe2} 
i\frac{d}{dt}c(k;t)=\int_{-\infty}^{\infty}dk' h(k,k';t) c(k';t).
\end{equation} 
Outgoing boundary conditions mean that one solves Eq.~(\ref{e:schroe2})
restricted to $k>0$:
\begin{equation}\label{e:schroe3} 
i\frac{d}{dt}c^+(k;t)=\int_{0}^{\infty}dk' h(k,k';t) c^+(k';t).
\end{equation} 
The time dependent wave function with outgoing boundary conditions
is then
\begin{equation}\label{e:expan3} 
\Psi^+(r;t):=\int_{0}^{\infty}dk' c^+(k';t)\Phi_{k'}(r)
\end{equation} 

In order to relate $\Psi(r,t)$ to the complex scaled wave function we
must assume that $\Psi(r,t)$ and its time--derivative
$d\Psi(r,t)/dt$ are analytic functions of $r$ for any analytic initial
state $\Psi(r,t=0)$.
This assumption is non--trivial: for example, it is known to be
valid for the Hamiltonian of the field--free hydrogen atom, while
it is obviously violated
for potentials that are non--differentiable at any point other
than $r=0$.
Under this assumption the analytically continued wave function
$\Psi(\eta r;t),\,Im(\eta)>0$ solves the ``complex scaled''
Schr\"odinger equation
\begin{equation}\label{e:complex1} 
i\frac{d}{dt}\Psi(\eta r;t)=H(\eta r;t)\Psi(\eta r;t),
\end{equation} 
which is the analogue of equation (\ref{e:schrcomp})
for a single radial coordinate.
If we further assume that the expansion functions
$\Phi_k$ are analytic, the expansion coefficients $c_{\eta}$
\begin{equation} 
\Psi(\eta r;t)=\int_{-\infty}^{\infty} dk\,c_{\eta}(k,t)\Phi_k(\eta r),
\end{equation} 
do not depend on $\eta$
\begin{equation} 
c_{\eta}(k,t)\equiv c(k,t).
\end{equation} 
From that it follows that also the kernel of the
time--integration $h(k,k';t)$ does not depend on $\eta$.
We see that the two equations (\ref{e:schroe1}) and (\ref{e:complex1})
describe exactly the same dynamics.
The important difference is that due to the asymptotic
behavior of the expansion functions $\Phi_k(\eta r)\sim\exp(ik\eta r)$
in Eq.~(\ref{e:complex1})
we can distinguish $k>0$ from $k<0$ by the norm:
``ingoing waves'' $k<0$  grow exponentially, while ``outgoing waves''
$k>0$ become square--integrable.
Because of the exponential divergence any function
\begin{equation} 
\int dk\, a(k)\Phi_{ k}(\eta r)
\end{equation} 
will diverge, if $\int_{-\infty}^{0}dk|a(k)|^2>0$.
Consequently, if $\Psi(\eta r;t)$ contains ingoing waves,
it will not be square integrable.
To obtain a solution with outgoing  waves only,
we solve the differential
equation (\ref{e:complex1}) restricted to the space of
square integrable functions $||\Psi^+(\eta r;t)||<\infty$.
Since $h(k,k';t)$ does not depend on $\eta$ the coefficients of
the expansion
\begin{equation} 
\Psi^+(\eta r;t)=\int_{0}^{\infty}dk' c^+(k';t)\Phi_{ k'}(\eta r)
\end{equation} 
are the same as in Eq.~(\ref{e:expan3}) and the outgoing wave
solution is obtained by substituting $\eta r$ with $r$.

As initial condition for Eq.~(\ref{e:complex1}) we use
a field free bound state. For the class of ``dilation analytic''
\cite{simon82}
potentials, which include the Coulomb potential, the bound state
functions are known to be analytic functions of $r$. This is trivial
to verify for the complex scaled Hamiltonian of the two--body
Coulomb problem
\begin{equation} 
\left(-\frac{1}{2\eta^2}\Delta-\frac{Z}{\eta r}\right)\Phi_i(\eta r)
=E_i\Phi_i(\eta r),
\end{equation} 
where $\Phi_i(r)$ is a hydrogenic bound state function.
For computation it is useful to keep in mind that,
for example, the radial ground state eigenfunction $\eta r\exp(-\eta r)$
becomes increasingly oscillatory with increasing
$Im(\eta)$, which is  opposite to the outgoing waves, where larger
$Im(\eta)$ causes stronger damping. Oscillatory functions are generally
more difficult to represent numerically and the choice of $\eta=e^{i 
\theta}$
will depend on whether the outgoing wave or the bound state
part of $\Psi_{\eta}$ are more important.

\newpage
\begin{table}
\caption{\label{t:basisl2}
Basis set for $L=2$ used in time propagation. ``Size'' denotes
the number of basis functions.
For the time--propagation near--linearly dependent vectors
are removed.
}
\begin{tabular}{cccccccr}
$ l$&$ \alpha_s$&$ \beta_s$&$k_s$&$m_s$&$n_s$&$p_s$&size \\
\hline
 0&   -2.000&   -0.333&    3& 9 & 2 &    9&\\
  &   -2.000&   -0.250&    3& 9 & 2 &    9&\\
  &   -2.000&   -0.200&    4& 9 & 2 &    9& 221\\
\hline
 1&  -1.400 &  -1.400 &    6&  6&  2&    6&\\
  &  -0.666 &  -0.666 &    6&  6&  2&    6& 56\\
\hline
 2&  -2.900 &  -2.900 &    1&  1&  1&    1&\\
  &  -1.400 &  -1.400 &    6&  6&  2&    6&\\
  &  -1.000 &  -1.000 &    6&  6&  2&    6&\\
  &  -0.666 &  -0.666 &    6&  6&  2&    6&  139\\
\hline
total &&&&&&&416 \\
\multicolumn{7}{l}
{total in time--propagation}& 318\\
\end{tabular}
\end{table}
\begin{table}
\caption{\label{t:energies1}
Bound state energies of $He$: (a) calculated with the basis used
for time propagation,  (b) literature values.}
\begin{tabular}{llc}
\multicolumn{1}{c}{(a)}&
\multicolumn{1}{c}{(b)}&
Ref.\\
\multicolumn{2}{c}{$L=0$}\\
-2.903724377&-2.903724392 &\cite{pekeris} \\
-2.14597404&-2.145974037  &\cite{pekeris}\\
-2.06127198&-2.0612719   &\cite{pekeris}\\
-2.0335877&-2.033586   &\cite{pekeris}\\
\multicolumn{2}{c}{$L=1$}\\
-2.123843088&-2.12384308&\cite{pekeris}\\
-2.05514636&-2.05514637&\cite{pekeris}\\
-2.0310696&-2.0310696&\cite{kono84}\\
-2.0199059&-2.0199059&\cite{kono84}\\
\multicolumn{2}{c}{$L=2$}\\
-2.055620727&-2.05562071&\cite{kono84}\\
-2.0312798&-2.0312798&\cite{kono84}\\
-2.0200158&-2.0200158&\cite{kono84}\\
-2.0138989&-2.0138981&\cite{kono84}\\
\multicolumn{2}{c}{$L=3$}\\
-2.0312551444&-2.03125514439&\cite{sims88}\\
-2.0200029370&-2.02000293714&\cite{sims88}\\
-2.0138906837&-2.01389068381&\cite{sims88}\\
-2.010205246& -2.01020524808&\cite{sims88}\\
\multicolumn{1}{c}{(a)}    & \multicolumn{1}{c}{(a)}\\
\multicolumn{1}{c}{$L=4$}  & \multicolumn{1}{c}{$L=6$}\\
-2.0200007108  		   & -2.0102041204\\
-2.0138893453    	   & -2.0078125284\\
-2.0102043836  		   & -2.0061728509\\
\multicolumn{1}{c}{$L=5$}  & \multicolumn{1}{c}{$L=7$}\\
-2.0138890346  		   & -2.0078125124\\
-2.0102041827  		   & -2.0061728489\\
-2.0078125737  		   & -2.0049999968\\
\end{tabular}
\end{table}

\begin{table}
\caption{\label{t:energies2}
Doubly excited states of $H\!e$ for $L=0$ through 3. (a) values
obtained with the basis used in the time
propagation, (b) literature values from references [29] ($L \leq 2$)
and [28] ($L>2$).
}
\begin{tabular}{cllll}
$(n_1,n_2)$&
\multicolumn{2}{c}{(a)}&
\multicolumn{2}{c}{(b)}\\
\multicolumn{5}{c}{$L=0$} \\
(2,2)&-0.777879&$4.55 \times10^{-3}$&-0.777868 &$4.53 \times10^{-3}$ \\
(2,2)&-0.621926&$2.156\times10^{-4}$&-0.6219275&$2.156\times10^{-4}$ \\
(2,3)&-0.589892&$1.37 \times10^{-3}$&-0.589895 &$ 1.35\times10^{-3}$ \\
(2,3)&-0.548085&$6.8  \times10^{-5}$&-0.5480855&$7.8  \times10^{-5}$ \\
(3,3)&-0.353517&$2.98 \times10^{-3}$&-0.353537 &$3.004\times10^{-3}$ \\
(3,3)&-0.317511&$6.9  \times10^{-3}$&-0.317455 &$6.67 \times10^{-3}$ \\
\multicolumn{5}{c}{$L=1$} \\
(2,2)&-0.6931347&$1.366\times10^{-3}$&-0.6931349
&$1.3773 \times10^{-3}$\\
(2,3)&-0.5970738&$3.857\times10^{-6}$&-0.59707381
&$3.84399\times10^{-6}$\\
(2,3)&-0.5640865&$2.93 \times10^{-4}$&-0.56408514
&$3.01057\times10^{-4}$\\
(3,3)&-0.335611 &$6.92 \times10^{-3}$&-0.3356269
&$7.023  \times10^{-3}$\\
(3,3)&-0.2862   &$3.04 \times10^{-4}$&-0.28595074
&$3.409  \times10^{-5}$\\
(3,3)&-0.282855 &$1.63 \times10^{-3}$&-0.28282897
&$1.46208\times10^{-3}$\\
\multicolumn{5}{c}{$L=2$} \\
(2,2)&-0.701938&$2.360\times10^{-3}$&-0.7019457&$2.3622\times10^{-3}$\\
(2,3)&-0.56925 &$6.9  \times10^{-4}$&-0.569221 &$5.55  \times10^{-4}$\\
(2,3)&-0.55640 &$3.6  \times10^{-4}$&-0.5564303&$2.01  \times10^{-5}$\\
(3,3)&-0.34309 &$5.174\times10^{-3}$&-0.343173 &$5.155 \times10^{-3}$\\
(3,3)&-0.31545 &$4.14 \times10^{-3}$&-0.31553  &$4.305 \times10^{-3}$\\
\multicolumn{5}{c}{$L=3$} \\
(2,3)&-0.5582830 &$1.297\times10^{-5}$&-0.55828&$1.28\times10^{-5}$\\
(2,3)&-0.5322936 &$3.50 \times10^{-5}$&\\			
(3,3)&-0.3042474 &$3.24 \times10^{-3}$&-0.30424&$3.24 \times10^{-3}$\\
(3,3)&-0.2780025 &$9.58 \times10^{-5}$&\\
\end{tabular}
\end{table}

\begin{table}
\caption{\label{t:energies3}
Bound and doubly excited states of $H^-$ for $L=0$ through 3. (a) values
obtained with the basis used in the time
propagation, (b) values with a basis optimized for each state, and 
literature
values.
The values (b) are estimated to be converged to all digits given except 
for the last.
}
\begin{tabular}{llc}
Energy & Width & Reference\\
\multicolumn{2}{c}{$L=0$} \\
-0.52775101689& 0 & present, (a)\\
-0.5277510165443& 0 &\cite{drake88}\\
\\
-0.1487764& 1.7324$\times 10^{-3}$&present, (a)\\
-0.1487762& 1.7332$\times 10^{-3}$&present, (b)\\
-0.1487765& 1.731 $\times 10^{-3}$&\cite{ho81}\\
\\
-0.12605&  12    $\times 10^{-5}$&present, (a)\\
-0.1260199& 9.02 $\times 10^{-5}$&present, (b)\\
-0.12601965&8.985$\times 10^{-5}$&\cite{ho95}\\
\\
-0.069006& 1.4192  $\times 10^{-3}$&present, (a)\\
-0.069006 &1.4184 $\times 10^{-3}$&present, (b)\\
\\
-0.05615&  23      $\times 10^{-5}$&present, (a)\\
-0.0561434& 8.8   $\times 10^{-5}$&present, (b)\\
\\
\multicolumn{2}{c}{$L=1$} \\
-0.12604986&$1.36 \times10^{-6}$&present, (a)\\
-0.12604986&$1.36 \times10^{-6}$&present, (b)\\
-0.1260495  &$1.165\times10^{-6}$&\cite{ho95}\\
\\
$^*)$&&present, (a)\\
-0.1243856&$7.0\times10^{-4}$&present, (b)\\
-0.12436  &$6.9\times10^{-4}$&\cite{ho93}\\
\\
-0.062708&  $1.17  \times10^{-3}$&present, (a)\\
-0.062716  &$1.19  \times10^{-3}$&present, (b)\\
-0.06871675&$1.1914\times10^{-3}$&\cite{ho92}\\
\\
-0.058586& $<10^{-5}$&present, (a)\\
-0.0585718&$8.988 \times10^{-6}$&present, (b)\\
-0.05857181&$8.986 \times10^{-6}$&\cite{ho92}\\
\\
\multicolumn{2}{c}{$L=2$}\\
-0.127937 &3.19  $\times 10^{-4}$&present, (a)\\
-0.127937 &3.12  $\times 10^{-4}$&present, (b)\\
-0.12794175&3.1625$\times 10^{-4}$&\cite{ho95}\\
\\
-0.065954&  1.654  $\times 10^{-3}$&present, (a)\\
-0.0659531& 1.6576 $\times 10^{-3}$&present, (b)\\
-0.0659533 &1.6581 $\times 10^{-3}$&\cite{bhatia90}\\
\\
-0.056834& 2.8 $\times 10^{-4}$&present, (a)\\
-0.0568294 &2.5302 $\times 10^{-4}$&present, (b)\\
\\
\multicolumn{2}{c}{$L=3$} \\
-0.056564& 3.54 $\times 10^{-3}$&present, (a)\\
-0.05655875& 5.00 $\times 10^{-3}$&present, (b)\\
\end{tabular}
{\footnotesize $^*)$ state cannot be distinguished from surrounding
continuum at the complex
scaling angle of $\theta=0.22$
used in time propagation}
\end{table}

\begin{table}
\caption{\label{t:248}
Ionization yield for laser wave length $248\,nm$ and pulse
duration $T=40$ optical cycles
as a function of peak intensity $I$. The literature values
are obtained by converting the generalized cross sections from Table~2
in Ref.~[36] with the help of Eq.~(16) in that reference.
}
\begin{tabular}{lll}
\multicolumn{1}{c}{$I\,(W/cm^2)$}&
\multicolumn{1}{c}{present}&
\multicolumn{1}{c}{Ref.~\cite{charalambidis97}}\\
$2\times10^{14}$  &$7.06\times10^{-3}$ & $8.13\times10^{-3}$ \\
$2.5\times10^{14}$&0.00105  & 0.00148 \\
$5\times10^{14}$  &0.043   & 0.069 \\
$1\times10^{15}$  &0.18     & 0.33 \\
\end{tabular}
\end{table}

\begin{table}
\caption{\label{t:harm}
Relative peak heights of harmonic generation by  $\cos^2$--pulses
of T = 40 optical cycles and peak intensity $2\times10^{14}W/cm^2$
for three different fundamental frequencies $ \omega$
The accuracies are 10\% up to the 5th harmonic and of the
order 50\% for the 7th harmonic.
}
\begin{tabular}{ccccc}
&\multicolumn{4}{c}{Harmonic order}\\
$ \omega$ & 1  & 3 & 5 & 7\\
\hline
0.34&0.85&0.25& $4.8\times10^{-5}$& $4.4\times10^{-8}$\\
0.38&1&3.9 & $3.2\times10^{-4}$& $3.4\times10^{-7}$\\
0.42&1.46& 0.11 & $2.1\times10^{-6}$&  --- \\
\end{tabular}
\end{table}

\newpage



\end{document}